# Brillouin light scattering studies of planar metallic magnonic crystals

**G. Gubbiotti[1], S. Tacchi[1], M. Madami[1], G. Carlotti[2], A. O. Adeyeye[3] and M. Kostylev[4]**

[1] CNISM, Unità di Perugia- Dipartimento di Fisica, Via A. Pascoli, I-06123 Perugia, Italy
[2] CNISM, Unità di Perugia- Dipartimento di Fisica and Università di Perugia, Via A. Pascoli, I-06123 Perugia, Italy
[3] Department of Electrical and Computer Engineering, National University of Singapore 117576, Singapore
[4] School of Physics M013, University of Western Australia, 35 Stirling Hwy, 6009 Western Australia, Australia

E-mail: gubbiotti@fisica.unipg.it

**Abstract.** The application of Brillouin light scattering to the study of the spin-wave spectrum of one- and two-dimensional planar magnonic crystals consisting of arrays of interacting stripes, dots and antidots is reviewed. It is shown that the discrete set of allowed frequencies of an isolated nanoelement becomes a finite-width frequency band for an array of identical interacting elements. It is possible to tune the permitted and forbidden frequency bands, modifying the geometrical or the material magnetic parameters, as well as the external magnetic field. From a technological point of view, the accurate fabrication of planar magnonic crystals and a proper understanding of their magnetic excitation spectrum in the GHz range is oriented to the design of filters and waveguides for microwave communication systems.

1. **Introduction**

Similar to photons in photonic crystals,[1] the spectrum of spin excitations in materials with periodically modulated properties shows bands of allowed magnonic states, alternated with forbidden band gaps.[2,3,4,5,6,7] This constitutes a new class of artificial crystals, now known as magnonic crystals (MC), in which collective spin excitations rather than light are used to transmit and process information.[8,9,10] Since the wavelength of these excitations are shorter than those of light in the GHz range, MC offer better prospects for miniaturization at these frequencies with the advantage that frequency position and width of the band gap are tunable by the applied magnetic field.

A MC can be formed starting from uncoupled resonators and making them coupled by some interaction, such as dipolar or exchange magnetic coupling. Alternatively, one can artificially process a continuous medium, to make a periodical profile of magnetic properties such as the saturation magnetisation and the exchange constant. Examples of one-dimensional (1D) MC, where the collective behaviour is mediated by dynamical dipolar interaction, are multilayered magnetic structures consisting of alternated ferromagnetic layers.[4] It is also possible fabricate a "lateral" multilayer in the form of an array of closely spaced parallel magnetic stripes or an array of stripes of different magnetic materials in direct physical contact with each other (Cobalt and Permalloy, for example).[11,12] Since the two materials have different coercivity and saturation magnetisation, this can be used to open the road to programmable magnonic ground state and to the consequent field-controlled dynamic response. The advantage of such a kind of "continuous" over discrete MC is that exchange coupling at boundary regions takes place and the dynamical dipole coupling is maximised. As a result, spin oscillations can easily transfer from one region to another and therefore spin waves can propagate across its entire structure with considerable group velocities. Remarkably, 1D MC in the form of lateral multilayers have also been obtained through the artificial modulation of the magnetic properties of a continuous film by either laser annealing or ion implantation.[13,14,15] More recently, Ki-Suk Lee an co-authors performed micromagnetic simulations on a novel 1D waveguide consisting of a Permalloy nanostripe with periodic modulation of its width.[16] Their predictions about allowed and forbidden bands of propagating dipole-exchange spin waves in such a system have been just confirmed by spatially–resolved Brillouin light scattering experiments in microsized modulating stripes.[17]

Similar to the case of 1D structure, also two-dimensional (2D) artificial magnonic crystals can be fabricated in the form of ordered arrays of either closely packed magnetic dots (coupled by dipolar interaction) or ferromagnetic antidots, i.e. a periodic array of holes drilled into a continuous magnetic film. In these systems the properties of collective spin wave can be controlled by changing the dots (holes) shape, dimension and symmetry arrangement in the array. The remarkable difference in the case of antidot samples is due to the presence of continuous portion of the magnetic film which makes possible the propagation of guided waves along certain particular directions.

The purpose of the present paper is to review the recent experimental observation of collective modes in 1D and 2D MC by the Brillouin light scattering (BLS) technique, which has proved to be a very powerful tool for the investigation of magnetisation dynamics in such structures. In this respect, BLS has some advantages with respect to other experimental methods, such as ferromagnetic resonance (FMR) with vector network analyzer (VNA)[18] and time resolved scanning Kerr microscopy (TRSKM),[19] usually employed to probe magnetisation dynamics in nanostructures. First of all, thanks to the wave vector conservation in the magnon-photon interaction, one has the possibility to measure the dispersion relation (frequency vs. wave vector) of the collective spin excitations, provided that the periodicity of the MC is such that the Brillouin zone (BZ) boundary lies in the wave vector range accessible in conventional BLS experiment. Furthermore, it has been recently shown that using a large-aperture objective, BLS can be used as scanning probe technique, permitting the map-out of the spatial distribution of magnetic normal modes with a lateral resolution of a few hundreds of nanometers.[20,21,22,23]

This paper has the following organization. Section 2 is dedicated to deep ultraviolet lithography technique and addresses its capability to obtain well-controlled nanostructures. In Section 3, the analytical theory used to calculate the spectrum of collective excitations in quasi-1D plane MC consisting of array of longitudinally magnetised stripes is described. In Sections 4.1 and 4.2 we review the experimental observation by BLS of collective spin excitations in arrays of non-contacting and contacting stripes, respectively, while Sections 5.1 and 5.2 contain results for 2D MC made by antidots and dots arrays. Eventually, Section 6 presents the conclusions and outlines some perspective in the field of MC.

**2. Fabrication and experiment 1D dense stripes array**

The fabrication of 1D and 2D magnonic crystals is based on recent abilities to produce dense arrays of magnetic elements arranged with sub-micrometer precision and with a very narrow distribution of shapes, sizes and distances. The statistical variations of these parameters can often hide, at least partially, the effect of magnetostatic and magnetodynamic interactions in arrays of thousands elements. Conventional nanofabrication techniques that are used in the microelectronic industry are not always compatible with magnetism because the process involves high temperature which will degrade the quality of the ferromagnetic films. It is also very difficult to use reactive ion beam etching to pattern magnetic films as it is not easy for the reactive gases to form volatile compounds when in contact with magnetic materials. Some of the key issues to be considered in the development of fabrication techniques for magnetic nanostructures are critical dimension control, resolution, size and shape homogeneity, patterned area and alignment accuracy. The nanofabrication methods used for synthesizing nanomagnets in recent years include electron beam lithography and lift-off processes, focused ion beam (FIB) etching, interferometric lithography, nanoimprint lithography, and anodic aluminum oxide membranes. For a general review of the various techniques for fabricating ordered magnetic nanostructures, the reader is referred to Refs. 24,25,26 and 27. Here we shortly describe the fabrication technique used at the University of Singapore to prepare most of the samples reviewed in this article, i.e. deep ultraviolet (DUV) lithography at 248 nm exposure wavelength. Alternating phase shift mask (PSM) was used to pattern large areas (typically 4x4mm$^2$) with either closely packed magnetic stripes or antidot arrays on commercially available silicon substrates,[28] with lateral dimensions much below the conventional resolution limit of optical lithography. One unique advantage of this technique is the fact that unlike e-beam lithography, thicker resists can be used to make high aspect ratio nanostructures. DUV lithography also has the ability to tune side wall profile by employing focus offset and resist processing temperatures. In addition, this nanofabrication technique is also compatible with conventional charge based complementary metal oxide semiconductor (CMOS) platform, thus enabling the integration of novel magneto-electronic devices. To create patterns in the resist, the substrate is coated with a 60 nm thick anti-reflective layer followed by 480 nm of positive DUV photoresist. A Nikon lithographic scanner with KrF excimer laser radiation is used for exposing the resist. To convert the resist patterns into ferromagnetic nanostructures, the Permalloy ($Ni_{80}Fe_{20}$) layer is deposited using physical vapour deposition techniques such as e-beam evaporation and sputtering on the resist patterns. The layer sitting on resist was lifted-off by dissolving the resist in solvent OK73 (trade name of the resist solvent). In the lift-off process it is crucial to have a clean break-off of the film at the pattern edges of the resist. To reduce step coverage, a collimating sample holder was designed. This holder restricts the incidence angle of the incoming material, thereby allowing only material in the path normal to the surface to reach the sample.[29] Lift-off was determined by the colour change of the patterned film and confirmed by examination under a SEM. With the special sample holder, the lift-off process was much easier and high aspect ratio patterned nanostructures with film thicknesses (> 120nm) were achieved. Note that the electron beam lithography and lift-off can also used to produce a lateral multilayer consisting of two families of stripes in direct physical contact with each-other, as made in Ref. 12. The first family of stripes (for example Permalloy stripes) is defined on polymethyl methacrylate (PMMA) resist followed by electron beam deposition and lift-off. For the fabrication of the second nanostripe array, another layer of PMMA resist is then deposited. At this point, high-resolution electron beam lithography with precise alignment can be used to define the position of the second nanostripe array. Subsequently, a second magnetic film (for example, Cobalt film) is deposited. The final structure thus consists of adjacent stripes in direct physical contact. (see Fig. 7)

**3. Theoretical model for the quasi-1D plane magnonic crystals**

Two main theoretical approaches have been used in the literature to describe the band structure of magnonic crystals, i.e. the plane wave method expansion and the dynamical matrix method. The former can be applied to periodical arrays in any dimension including periodic arrays of magnetic particles embedded in a non-magnetic medium or in a matrix made of a different ferromagnetic material.[30,31,32,33] The latter is an hybrid method[34,35] combining micromagnetic simulations for calculating the ground state and an eigenvalue/eigenvector approach, which requires the computation of a dynamical matrix, whose elements are related to the torque acting on the magnetisation in each cell. The eigenvalues are the frequencies of the magnonic modes of the system and the eigenvectors give spatial profiles of the modes. The elements of the dynamical matrix can be calculated analytically for any contribution to the total energy (magnetostatic, exchange, Zeeman, magnetic anisotropy) without introducing numerical approximations, while the final diagonalization of the dynamical matrix is done numerically. So far, this method has been applied mainly to single planar dots of different shapes.[23,35,36] One also has to note that different micromagnetic simulation packages, such as OOMMF,[37] can be adopted for studying coupled dynamics of nanoelements, following the space resolved dynamics of magnetisation in time domain and operating a Fourier transform to the frequency domain.[38,39,40] However, this latter method does not provide information on the frequency dispersion over the whole artificial Brillouin zone. If one restricts to the case of collective spin excitations in 1D-MC, it is possible to formulate analytical models which yield direct physical insight to the characteristics of collective excitations.[2-6,41] One of these analytical approaches is presented in the following to analyse the case of 1D *plane* MC consisting of dynamically coupled nanostripes of (quasi)-infinite length, magnetised along their length. Within this framework, it is quite straigthforward to derive a number of results which can easily illustrate the effect of dynamical coupling in systems of interacting stripes, as we do in the following. The starting point to solve this problem is the linearized Landau-Lifschitz equation of motion for magnetic moment

$$-i\omega \mathbf{m}(x,y,z) = -\gamma \left[ \left( \mathbf{M}(x) + \mathbf{m}(x,y,z) \right) \times \left( \mathbf{H} + \mathbf{h}(x,z) \right) \right] \quad (1)$$

In this equation γ is the gyromagnetic coefficient, **M** is the equilibrium magnetisation, with |M| equal to the film saturation magnetisation $M_s$. The equilibrium magnetisation is directed along the magnetic field **H** which is applied in the film plane. To be specific we will consider the situation which is usually implemented in Brillouin light scattering experiments: **M** and **H** are directed in the plane array along the stripe longitudinal axis *y*. In this magnetic ground state the stripes are completely saturated and there is no static demagnetising field (the stripes are of infinite length along *y*). The equilibrium magnetisation depends on the co-ordinate *x* which lies in the array plane and is perpendicular to the longitudinal axis *y*. This is the direction of array periodicity and thus |**M**| changes periodically. In particular, if an array of stripes separated by air nanogaps is considered |**M**| varies between $M_s$ in the stripes and zero in the gaps. As **M** and **H** are aligned along *y* the small linear dynamic magnetisation component **m** has one component in the array plane ($m_x$) and one perpendicular to the array plane ($m_z$). Its precession frequency is ω. This spatially inhomogeneous precessing magnetisation gives rise to an effective field **h** which in our case of free oscillations consists of an effective exchange field

$$\mathbf{h}_{exc} = \alpha \nabla^2 \mathbf{m} \quad (2)$$

where $\alpha$ the exchange constant, and of a dynamic dipole field $\mathbf{h}_d$ which is described by the magnetostatic equations

$$\begin{cases} \nabla \times \mathbf{h}_d(x,z) = 0 \\ \nabla \cdot \mathbf{h}_d(x,z) = -4\pi \nabla \cdot \mathbf{m}(x,z) \end{cases} \quad (3)$$

Three different methods were used to solve the system (1)-(3) for 1D plane magnonic crystals.[12,14,41] They mainly differ by the way how the magnetostatic equations (3) are solved. In Ref. 12 this system of equations is transformed into a 2D Poisson equation for the magnetostatic potential. In Ref. 14 a known simple solution in the Fourier space[42] is used and the system of equations (2-3) is transformed into an infinite system for spatial Fourier amplitudes of dynamic magnetisation. In this work we will follow the approach of the Green's function in the direct space described in Refs. 43 and 44. Importantly, all free approaches require numerical treatment of the problem in the last stage of solution, however the Green's function approach allows drawing some important qualitative conclusions based on approximate analytical formulas.

We assume that along the stripe the magnetisation dynamics and the effective field take the form of a propagating wave

$$\mathbf{m}(y,x,z) = \mathbf{m}_{k_y}(x,z)\exp(ik_y y),$$
$$\mathbf{h}(y,x,z) = \mathbf{h}_{k_y}(x,z)\exp(ik_y y). \qquad (4)$$

With this assumption one arrives at a Green's function for the dipole field. In the framework of the Green's function approach the dipole field is related to the magnetisation through an integral relation, as follows

$$\mathbf{h}_{dk_y}(x,z) = \int_S \hat{G}_{k_y}(x-x',z-z')\mathbf{m}_{k_y}(x',z')dx'dz' \qquad (5)$$

where $S$ is the cross-sectional area of the magnetic material. The Green's function $\hat{G}_{k_y}$ takes the form of an integral relation[41] which is not difficult to compute numerically. It can be easily expressed in terms of the zeroth-order modified Bessel function of the second kind $K_0$:

$$\hat{G}_{k_y,\alpha\alpha'}(x,z) = \frac{\partial}{\partial\alpha}\frac{\partial}{\partial\alpha'}K_0\left(k_y\sqrt{(x-x')^2+(z-z')^2}\right) \qquad (6)$$

where $\alpha$ and $\alpha'$ are $x$ or $z$ and $x'$ or $z'$, respectively. Using Bloch-Floquet theorem the solution along the periodicity direction takes the form of a Bloch wave. This solution reads[43]

$$\mathbf{m}_{k_y}(x,z) = \tilde{\mathbf{m}}_{k_y}(x,z)\exp(-ikx) \qquad (7)$$

where $\tilde{\mathbf{m}}_{k_y}(x,z) = \tilde{\mathbf{m}}_{k_y}(x,z+a)$ is a periodical function with the period equal to the quasi-crystal lattice constant $a$, and $k \equiv k_x$ is the Bloch wavenumber. A similar expression is valid for the effective field. Based on Eq. 7, the Green's function in the *reduced-zone* scheme reads:

$$\tilde{\mathbf{h}}_{dk_y,k}(x,z) = 4\pi \int_{a\times L} \tilde{G}_{k_y,k}(x-x',z-z')\tilde{\mathbf{m}}_{k_y,k}(x',z')dx'dz' \qquad (8)$$

with $\tilde{G}_{k_y,k}(x-x',z-z') = \sum_{n=-\infty}^{\infty}\hat{G}_{k_y}(x-x',z-z')\exp[ik(x-x'-na)]$, and with the integration over the cross-section $a\times L$ of just one structure period ($L$ is the stripe thickness along $z$).

On substitution of (8), (7), and (2) into the equation (1), this latter transforms into a self-consistent integro-differential equation which can be easily solved numerically. We used this way in Ref. 44 to obtain the complete quasi-2D treatment of the case of an array of stripes separated by air gaps. In this case the function $\mathbf{M}(x)$ in Eq.(1) is trivial and the integration in (8) reduces to integration over the cross-sectional area of one stripe. In contrast, for an array consisting of stripes with two different values of saturation magnetisation on a structure period,[12] one has to integrate over the whole period's cross-section.[41]

Importantly, in many cases the Green's function (6) can be considerably simplified without noticeable loss of accuracy.[43,44] As shown previously by Guslienko et al.,[45] one can use averages of the dynamic magnetisation and of the dipole field across the element thickness $L$ to obtain one-dimentional formulas which allow a simple analytical insight (see Appendix to Ref. [46] for description of the averaging procedure). This approach is valid for the stripes with a small aspect ratio $L/w<<1$.

For instance, for a collective mode propagating along the array periodicity direction ($k_y$=0), the Green's function $\hat{G}_k$ reduces to a simple equation first proposed by Guslienko and co-workers[45]

$$G^{zz}(s) = -G^{xx}(s) - \delta(s) = 1/(2\pi L)\ln[s/(s^2+L^2)] \quad (9)$$

where $s=x-x'$. Importantly, the rest of Green's function components vanishes in this approximation for $k_y$=0. As reported in Ref. 45, due to this simple form for the Green's functions with two components differing by the trivial term $\delta(s)$ only, solution of the system (1)-(3) is considerably simplified in the important case of an array of stripes all made from the same material and separated by air gaps. One does not need to seek for a solution which satisfies (1) and (3) simultaneously The magnetostatic equations reduce to a "self-consistent" eigenvalue problem for an integral operator

$$N_{zzk}\tilde{m}_{zk}(x) = -\int_{-w/2}^{w/2}\sum_{n=-\infty}^{\infty} G^{zz}(x-x'-na)\exp[ik(x-x'-na)]\tilde{m}_{zk}(x')dx' \quad (10)$$

where $w$ is the stripe width. Once the set of eigenvalues $N_{zzk_B}$ has been found, they can be substituted into (1). The physical meaning of $N_{zzk_B}$ is easily found from comparison of Eqs.(10) and (8). One sees that inside the stripes $\tilde{h}_{dzk_B} = -4\pi N_{zzk_B}\tilde{m}_{zk_B}(x')$, thus $N_{zzk_B}$ represents the effective out-of-plane demagnetising factor for the collective modes. The in-plane demagnetisation factor is also easy to find: $N_{xxk_B} = -1 - N_{zzk_B}$. The rest of the components of the tensor of the effective demagnetisation factors vanishes. Importantly, each particular eigenmode $n$ possess its own value of the demagnetisation factor which also depends on the Bloch wavenumber.
Neglecting the exchange interaction, which is appropriate for several lowest-order collective modes of the array, provided the stripe width is larger than 300 nm or so (for saturation magnetisation of Permalloy) Eq.(1) reduces to a simple algebraic equation for eigenfrequencies of collective modes $\omega_{nk_B}^2$ involving $N_{xxk_B}$:[43]

$$\omega_{nk}^2 = \gamma^2 H(H + 4\pi M_s) + \gamma^2(4\pi M_s)^2\left(N_{xxk} - N_{xxk}^2\right) \quad (11).$$

For homogeneous magnetisation precession in a continuous ferromagnetic film $N_{xxk_B}$ vanishes and one recovers the lowest possible frequency, i.e. the in-plane ferromagnetic resonance frequency. If the film is patterned in an array of stripes, in-plane confinement is introduced which results in a non-vanishing value for $N_{xxk_B}$ and in an increase of the eigenfrequency in agreement with Eq. 11. Here one has to recall that the approach of averaging over the film thickness is an approximation. Only $N_{zzk_B}$ larger than ½ (and $N_{xxk_B} < 1/2$) make sense.[41] The rest should be rejected. For the stripes with small aspect ratio all experimentally observable modes meet this criterion $N_{xxk_B} < 1/2$. From (10) one also sees that the eigenfunctions $\tilde{m}_{zk_B}(x)$ represent the eigen-distributions of dynamic magnetisation across the stripe width for different modes. Let us consider an array consisting of identical stripes, e.g. one from Ref. 47. Numerical solutions for Eq. (11) in this case are shown in Fig. 1 which demonstrates the fundamental collective mode of the array and the first higher-order mode.

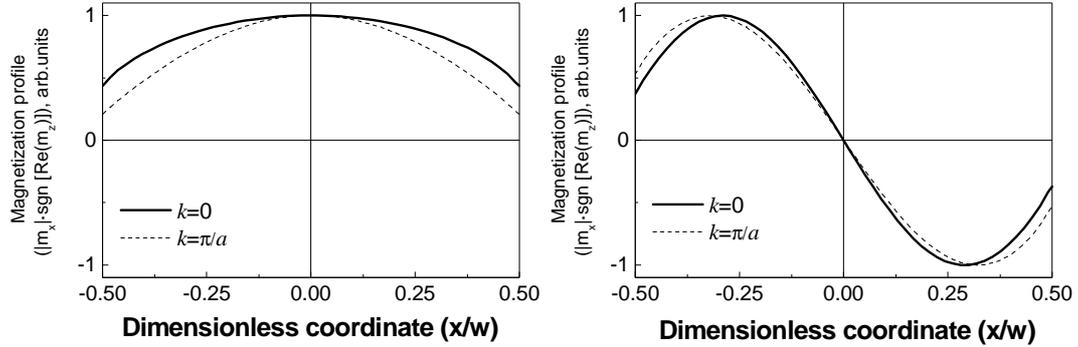

**Fig. 1** Comparison between the calculated profile of dynamic magnetisation for the lowest frequency mode (left) and the next one (right panel) for an array of stripes of widths $w$=175 nm, separation $\Delta$=35 nm, thicknes 20 nm, period $a$=210 nm. Solid line (dashed line) curves refer to the case of $k$=0 ($k=k_{BZ}=\pi/a$). The dimensionless coordinate in abscissa is normalized to the stripes width ($w$). Reprinted with permission from [47]. G. Gubbiotti, G. Carlotti, P. Vavassori, M. Kostylev, N. Singh, S. Goolaup, and A. O. Adeyeye 2005 *Phys. Rev. B* **72** 224413. © 2005, American Institute of Physics.

From Fig. 1 one sees that the mode profiles have the same shapes as the resonant modes for uncoupled (individual) stripes.[47] However, a closer analysis reveals that the profiles shapes are Bloch-wave number dependent and that the amplitude of dynamic magnetisation at the stripe edges which reflects the strength of the effective dipole pinning of magnetisation[47] also differs from the case of the individual stripes. Let us consider the first the fundamental mode. In the middle of the first Brillouin zone (BZ) $k$=0, the mode amplitude at the edges is larger (the effective pinning is smaller) for the dipole-coupled array of stripes than for uncoupled stripes. On the contrary, for the edge of the first BZ ($k=k_{BZ}=\pi/a$) the effective pinning is larger for the coupled array that for the individual stripes.

For the first higher-order mode which has one node of distribution of dynamic magnetisation across the stripe width an opposite behaviour can be observed: the effective pinning of edge magnetisation is larger for $k$=0 than for the individual stripes while it is smaller for $k=k_{BZ}$. This difference is related to the amount of the dipole energy contained in different modes and thus to the mode frequencies. To get an idea of the strength of dipole coupling let us consider the stripe dipolar field inside and outside the stripe. Inside each stripe the dipole field consists of the field produced by the stripe itself and of the dipole fields from the neighbours. The field from the neighbours determines the strength of dipole coupling. Consider the dipole field of an individual stripe in its vicinity. It can be calculated by solving the eigenvalue/eigenfunction problem for the integral operator $G^{zz}(s)$ in Eq.(9).[45] Once the eigenfunctions are found, one uses Eq.(5) to calculate the dipole field created by different modes outside the stripes. The result of such calculation is shown in Fig. 2. From this figure one sees that the dipole field of the fundamental mode (index "0") is considerably more far-reaching than for any other mode. Thus one has to expect the largest dipole coupling strength for the fundamental mode. Indeed, the effective demagnetisation factor for this mode and thus the mode eigenfrequency differ the most from the respective mode of an individual stripes. The effect of dipole coupling on the other modes for the array is almost negligible and the frequencies are practically the same as for the respective modes on the individual stripes.

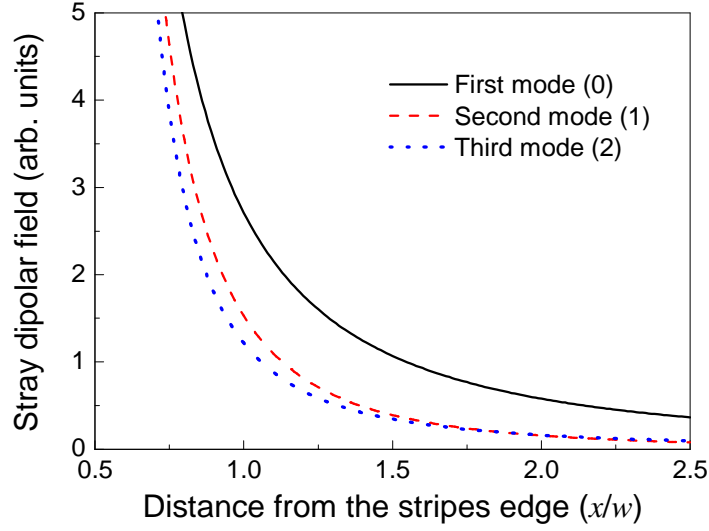

**Fig. 2** In-plane component of the dynamical stray field created by different magnetic normal modes in a longitudinally magnetised stripe at a distance $x$ from its edge (distance is normalized with respect to stripe's width). Numbers ($n$) labels the nodal planes of different modes.

Not only the difference in the frequency from individual stripes is important, even more important is how the frequency varies across the first BZ. To get an idea of this it is sufficient to consider only two points inside the zone $k=0$ and $k=k_{BZ}$. For these points Eq.(10) reduces to a simpler formula: [43]

$$\tilde{h}_{dzk}(x) = -4\pi N_{zzk}\tilde{m}_{zk}(x) = 4\pi \int_{-w/2}^{w/2} \sum_{n=-\infty}^{\infty} C_{nk} G^{zz}(x-x'-na)\tilde{m}_{zk}(x')dx' \quad (12),$$

where $C_n = 1$ for $k=0$ and $C_n = (-1)^n$ for $k=k_{BZ}$. Importantly, the in-plane component of the dipole field $\tilde{h}_{dxk_B}(x)$ changes its sign on the stripe edge $x=w/2$. It is seen from the magnetostatic boundary condition $\tilde{h}_{dxk_B}(x=w/2-0) + 4\pi\tilde{m}_{dxk_B} = \tilde{h}_{dxk_B}(x=w/2+0)$, keeping in mind that inside the stripes the dipole field should be anti-aligned to the dynamic magnetisation. Then one sees that the dipole field induced by the neighbours sums up in anti-phase with the stripe's own dipole field for the middle of the first BZ. As follows from (12), this reduces the value for $N_{xxk_B}$ and the total dipole energy. Thus the eigenfreqency is minimized (recall that $N_{xxk_B} < 1/2$ always). This decrease in the total dipole energy reduces the effective dipole pinning.

On the contrary, at the edge of the first BZ the dynamic magnetisation in the adjacent stripes oscillates in anti-phase ($C_n = (-1)^n$) this results in stripe's own dipole field adding up in –phase with the fields from the nearest neighbours. This increases $N_{xxk_B}$ and the frequency goes up compared to the individual stripes. Importantly, this $N_{xxk_B}$ value is the maximum possible for the fundamental mode. Thus we see that the fundamental mode acquires its minimum frequency at $k=0$ and its maximum frequency at $k=k_{BZ}=\pi/a$. The frequency band for the fundamental magnonic mode will obviously decrease with increase in the width of the air gap between the stripes, as the amplitude of dipole field decreases with distance (see Fig. 2). A good example of the frequency width for the fundamental magnonic band is given in Fig. 6 of Ref. 43.

In the same way, the behaviour of the first higher order collective mode can be explained. First, one has to take into consideration that the dipole field of this mode outside the stripes is considerably weaker due to spatially inhomogenous precession (see Fig. 1). Furthermore, it is quickly decreasing with the distance from the stripe edge. Therefore, the half of the stripe width which is closer to a particular neighbour experiences larger coupling to this neighbour than the other half. Consider two dipole-coupled stripes in the chain. In the middle of the first BZ precession in all dipole coupled stripes is in-phase. Thus, the same profile of magnetisation as in Fig. 1 is valid for both stripes. One sees that the negative half-period of the standing wave in each stripe is adjacent to the positive half-period of the standing wave in its right-hand side neighbour. The same consideration as for the fundamental mode applies for the phase of the contribution to the dipole field coming from the nearest neighbours. The phase of this contribution is in phase with the phase of precession inside the neighbour close to its edge. Thus, for $k=0$ the stripe's own dipole field is in-phase with the contribution from the closest neighbour and both fields are anti-aligned with dynamic magnetisation. The dipole energy is maximised and one retrieves the maximum frequency for this mode. On the contrary, for $k=k_{BZ}$ the magnetisation precession in two adjacent stripes near the edges of the same gap is in phase and the stripe's own dipole field is in anti-phase to the contribution from the nearest neighbour. The frequency for this mode is minimised. Thus the dispersion of the first higher-order collective mode should be negative in the first BZ between $k=0$ and $k=k_{BZ}$.

**4.1 Experimental study of 1D magnonic crystals: Arrays of closely packed stripes**

Here follows a list of stripes arrays fabricated using the method discussed in Section 2 and whose properties will be reviewed here.

#1 Homogeneous nanostripes array of width $w$=175 nm, separation $\Delta$=35 nm, thickness 20 nm [period $a$=210 nm, BZ boundary ($k_{BZ}=\pi/a = 1.5\times10^5$ rad/cm)]

#2 Homogeneous nanostripes array of width $w$=350 nm, separation $\Delta$=55 nm, thickness 30 nm [period $a$= 405 nm, BZ boundary ($k_{BZ}=\pi/a=0.78\times10^5$ rad/cm)]

#3 Alternated nanostripes array consists of nanostripes of widths $w_1$=350 nm and $w_2$=500 m, separation $\Delta$=55 nm, thickness 30 nm. [period $a$= 960 nm, BZ boundary ($k_{BZ}=\pi/a=0.33\times10^5$ rad/cm)]

Fig. 3 shows the scanning electron micrograph (SEM) images of the stripe arrays and reveal well defined stripes with uniform spacing and width.

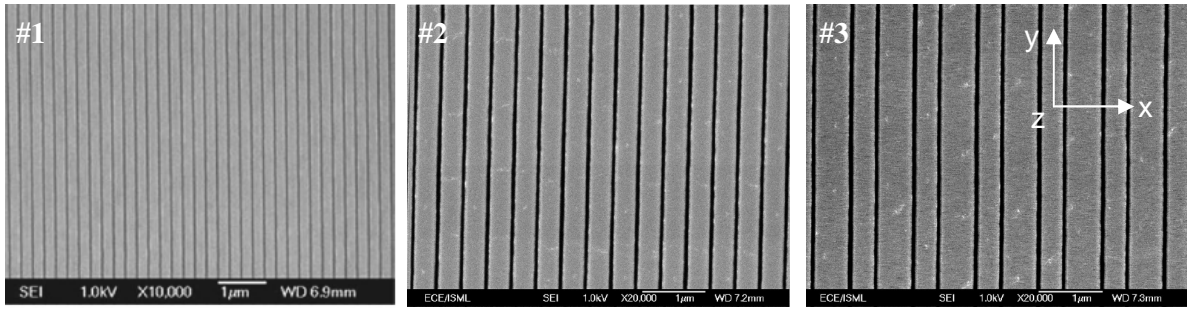

**Fig. 3** SEM micrographs of the stripes arrays. The Cartesian coordinate system used for the calculation is also reported. Reprinted with permission from [3]. G. Gubbiotti, S. Tacchi, G. Carlotti, N. Singh, S. Goolaup, A. O. Adeyeye, and M. Kostylev 2007 *Appl. Phys. Lett.* **90** 092503. © 2007, American Institute of Physics. Reprinted with permission from [47]. G. Gubbiotti, G. Carlotti, P. Vavassori, M. Kostylev, N. Singh, S. Goolaup, and A. O. Adeyeye 2005 *Phys. Rev. B* **72** 224413. © 2005, American Institute of Physics.

Figure 4 shows measured and calculated dispersion curves for all the investigated stripe arrays. Brillouin light scattering experiments were carried out by the Perugia group and spectra were measured at room temperature with 220 mW of *p*-polarized monochromatic light from a solid state laser $\lambda$=532 nm. Incident light was focused onto the sample surface using a camera objective of numerical aperture 2 and focal length 50 mm while the s-polarized backscattered light was frequency analyzed by a (3+3) tandem Fabry-Pérot interferometer. Spectra were recorded in the Voigt configuration where the external magnetic field H was parallel to both the sample plane and the stripes length (y direction) while the in-plane wave vector transferred in the scattering process was perpendicular to H (x direction). The sample was mounted on a goniometer that allowed us to vary the incidence angle of light θ between 0° and 70°. The angle θ is linked to the probed spin-wave wave vector magnitude by the relation $k=(4\pi/\lambda)\sin\theta$.

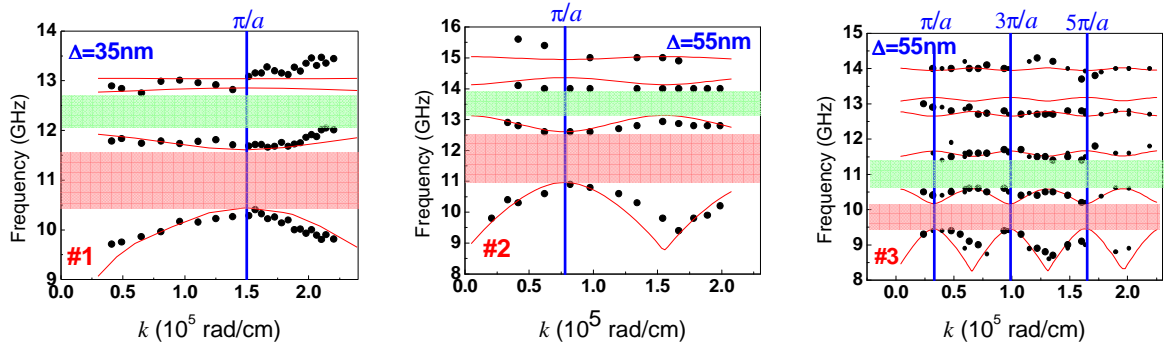

**Fig. 4** Experimental and calculated spin-wave frequency dispersion for an easy axis applied magnetic field $H$=500 Oe for stripes arrays #1, #2 and #3 described in the text. Vertical lines indicate the edges of the Brillouin zone ($k=n\pi/a$) while colored areas indicate the frequency band gap. Reprinted with permission from [47]. G. Gubbiotti, G. Carlotti, P. Vavassori, M. Kostylev, N. Singh, S. Goolaup, and A. O. Adeyeye 2005 *Phys. Rev. B* **72** 224413.© 2005, American Institute of Physics. Reprinted with permission from [3]. G. Gubbiotti, S. Tacchi, G. Carlotti, N. Singh, S. Goolaup, A. O. Adeyeye, and M. Kostylev 2007 *Appl. Phys. Lett.* **90** 092503. © 2007, American Institute of Physics.

BLS measurements as a function of the transferred in-plane wave vector reveal that magnetic modes have an oscillating dispersive character with the appearance of Brillouin zones determined by the artificial periodicity of the stripes array. Because of the different quasi-crystal lattice period $a$ samples possess different width of the 1st BZ. The collective mode dispersion is periodical (period of 2 $k_{BZ}$) and successive modes oscillate in anti-phase. The oscillation amplitude (amplitude of the magnonic band) is more pronounced for the fundamental collective mode (the lowest frequency mode resonating along the stripes width) and decreases for the highest modes. This is in agreement with our prediction above, which was based on the modes' dipole (stray) fields outside stripes. As also discussed above, the dispersion slope for the fundamental mode between $k=0$ and $k=k_{BZ}$ is positive while for the first higher-order mode (which is next in frequency to the fundamental one) it is negative. Each mode exists in its frequency range separated from the neighbouring by a prohibited zone. This is different from the case of non-interacting stripes where, as a consequence of lateral confinement, quantized spin wave have been observed. These quantized modes have a stationary character and are dispersion-less, i.e. their frequency does not change over the whole range of wave vector investigated.[47,48]

Going from sample #1 to #3 the artificial periodicity increases with the consequent reduction of the Brillouin zone width. This means that for sample 1 just one complete frequency oscillation is visible while for sample #3 oscillation over the fourth BZ is observed. One also sees a good agreement of experimental results with the model described above. However , one has to note, that in all figures the frequency width of the fundamental magnonic band $\omega(k_{BZ})- \omega(k=0)$ is larger than the experimentally measured one. Importantly, all our other measurements systematically showed smaller bandwidths for the fundamental mode than calculated.[3,44,47] Interestingly, that this tendency is also seen in Ref. 12 where a 1D magnonic crystal of a different type was studied and where one used a different approach to calculate the dispersion.

To this respect, one has to notice that there is no adjustable parameter in the model, which can reduce the mode bandwidth without affecting the frequency positions for all higher-order modes. The only model's parameter which affects the magnonic band bandwidth without noticeably affecting the positions of the higher-order non-dispersive modes is the strength of the dipolar coupling. This latter quantity is completely determined by the array geometry. Therefore, to account for this systematic discrepancy we have to assume that the geometry is not perfect and the reduction in the bandwidth is connected to reduction in coupling strength due to small imperfections of a real quasi-crystal. Possible structural imperfections which can affect the bandwidth are small variations in the widths of the gaps from period to period and along the same gap (along $y$) and slight variations in stripes' width and in stripes' positions on individual structure periods.

On the other hand, it was found that one can easily fit the mode bandwidths with the theoretical formulas, provided one decreases the number of the nearest neighbours $N$ whose dipole field contributions are accounted in the calculation of the collective dipole field. This is given by the summation range over $n$ in Eq.(10). In the ideal case all stripes on the infinite chain contribute to the collective demagnetising factor ($N=\infty$). To fit the experimental data one takes into account a smaller number of the nearest neighbours $N_c$ ($n=-N_c, -N_c+1\ldots N_c$ in Eq.(10)). This reduces the dipole field strength which in its turn reduces the frequency bandwidths for all collective modes. This fact suggests that $2N_c+1$ is the number of nearest neighbours whose precession phases are correlated due to dipole coupling. Thus $(2N_c+1)a$ can be considered as a correlation (coherence) length for a mode which is smaller than infinity due to the quasi-crystal imperfections. This correction to the model (10) is purely empirical. Therefore a question arises, why this model for the reduction of the value for collective $N_{zzk_B}$ is preferred? Alternatively, what is the reason for summing up contributions from all nearest neighbours with $|n|\leq N_c$ with equal weights $W_n=1$ equal to 1 and from all other stripes ($|n|>N_c$) with a zero weight ($W_n=0$)? While answering this question, we find that a model in which contributions from the neighbours are weighted according to the following expression $W_n=\exp(-n/N_c)$ gives the same good fits for the experimental data (certainly, $N_c$ values which result in an optimal fit are different from those obtained in the approach of the equal weights).

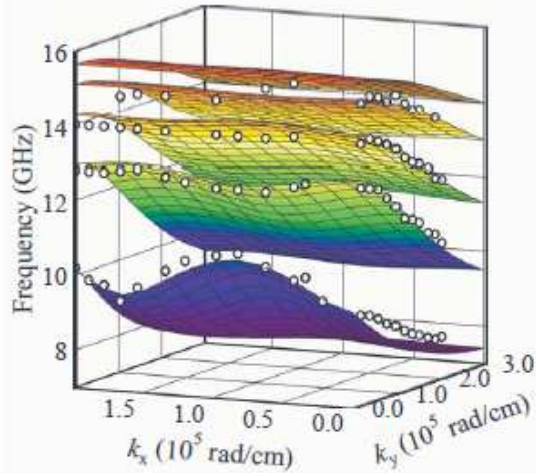

**FIG. 5.** Spectrum of collective modes in the two-dimensional space $k_x=k, k_y$. Surfaces are calculations and dots are experimental points measured as a function exchanged wave vector along the $\varphi=0°$ (parallel to stripes length) and $\varphi=90°$ (perpendicular to stripes length) directions. Reprinted with permission from [44]. M. Kostylev, P. Schrader, and R. L. Stamps, G. Gubbiotti and G. Carlotti, A. O. Adeyeye, S. Goolaup and N. Singh 2008 *Appl. Phys. Lett.* **92** 132504 © 2008, American Institute of Physics.

This latter approach of weighted neighbours' contributions was used to fit the experimental data for a number of magnonic crystals, including ones studied in Refs. 44 and 57. For all studied arrays of stripes the coherence lengths as extracted from fits to experimental data is several structure periods. In particular, in Fig. 5 the agreement theory with experiment is obtained for $N_c$=2. This number means that coupling of second neighbours is by (exp(1)) times smaller that in the ideal case. Figure 5 was taken from Ref. [44] where we measured the spin wave frequency dispersion both along the stripes length ($\varphi$ =0°) and perpendicular ($\varphi$ =90°) to it. Results are presented in Fig. 5. For $\varphi$ =90°( BLS transferred wavenumber swept along the stripes width), the collective modes represent coupled resonances of Damon–Eshbach magnetostatic surface wave. In the case of dipole coupling of such resonances a Bloch wave is formed, characterized by Brillouin zones induced by the artificial periodicity. The lowest resonance has a quasi homogeneous distribution of the dynamic magnetisation across the stripe width, as discussed above, while the higher order collective modes have either a positive or a negative slope depending on the specific parity of the dynamic magnetisation profile across the stripe widths. For $\varphi$= 0° (BLS transferred wave number swept along the stripes length) the dispersion becomes aperiodical (monotonic). This is what one has to expect, since our crystal is quasi-1D, and the direction $\varphi$= 0° is perpendicular to the periodicity axis. The waves propagating along the stripes are of backward volume wave (BVMSW) type.[49] Their dispersion is modified due to dipole coupling between the stripes which decreases dispersion slope with respect to the dispersion of guided BVMSW on uncoupled stripes.[46] In this work we also went beyond the conventional BLS scattering geometry (MSSW ($\varphi$= 90°) and BVMSW($\varphi$= 0°)) and observed variation in the mode frequency as a function of the waves propagation angle $\varphi$ with respect the applied field direction. Rotation of the vector $k$, with fixed modulus, from $\varphi = 0°$ to $\varphi = 90°$, corresponds to transition from a modified BVMSW geometry to a modified MSSW geometry. BLS spectra were measured by keeping the incidence angle of light against the sample normal fixed ($\theta$=30° which gives $\left(k_x^2 + k_y^2\right)^{1/2}$=1.18·10$^5$ rad/cm). A fixed external magnetic field H=1.0 kOe was applied parallel to the stripes length. In our measurements and calculations we found that the frequencies for all the modes decrease as the angle $\varphi$ varies from 90° to 0°(see Fig.6, right).

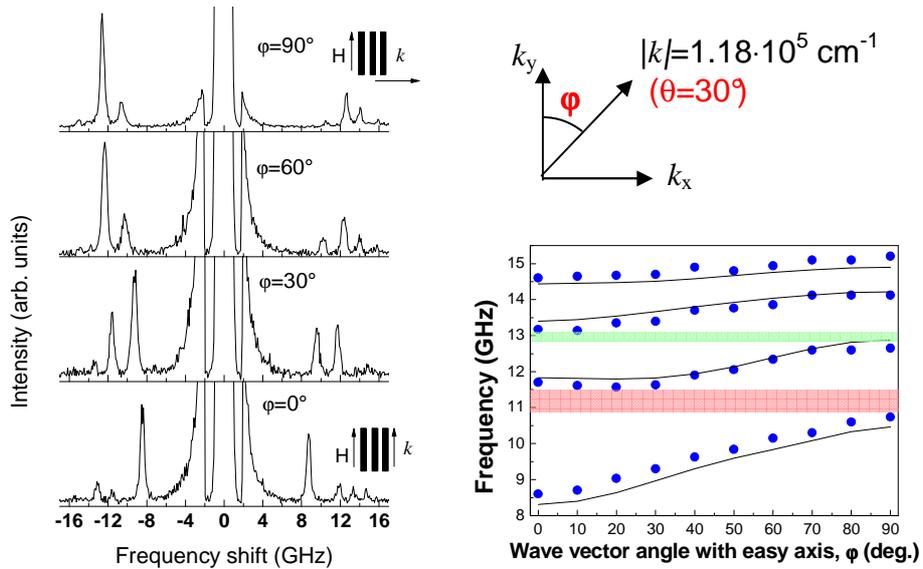

**FIG. 6** (Left panel) Sequence of Brillouin light scattering spectra measured for H=1.0 kOe at different in-plane angles $\varphi$ between the stripes length ($y$ axis) along which H is applied and the scattering plane. Insets show the scattering geometries for $\varphi$ =0° and $\varphi$=90°. (Right panel) Dependence of collective mode frequency on the wave propagation angle $\varphi$. Dots: experiment, solid lines: numerical calculation. Band gaps are also shown by colored areas. Reprinted with permission from [44]. M. Kostylev, P. Schrader, and R. L. Stamps, G. Gubbiotti and G.

Fig. 6 contains a sequence of measured BLS spectra for different values of the in-plane angle φ. A closer inspection of the measured spectra reported in Fig. 6, reveals that the largest intensity peak is the lowest frequency modes for φ up to 30° while it is the second for φ larger than 30°. This is ascribed to the fact that increasing the φ value determines an increase of the transverse wave vector component ($k=k_x$) and consequently a difference in the collective modes probed from the first to the second Brillouin zone. A recent theoretical paper[50] discusses peculiarities of Brillouin light scattering response from magnonic modes on periodic arrays of stripes. An interested reader may find discussion of influence of parameters of experiment and magnetic properties of the sample (such as magnetic damping) on BLS intensities. With the experiment showed in Fig. 6, we demonstrated that planar periodical structures of parallel dipole-coupled magnetic stripes support propagation of Bloch waves at any angles with the major structure axis. Due to lateral confinement within the stripe width, the eigenwaves spectrum has a large number of modes, while the artificial periodicity induces the presence of Brillouin zones and consequent periodicity in *k* space. Noticeably, the frequency gaps in the spectrum are partial, i.e., the stop bands for propagation along the major stripe axis and the frequency pass bands for propagation perpendicular to it overlap. This is connected to the fact that the artificial magnonic crystal studied here is 1D while the occurrence of a full band gap would require at least two-dimensional periodicity.

As a general comment on the width of allowed and forbidden spin wave bands, we notice that in the case of 1D array of closely-packed stripes, these are much smaller (1-2 GHz) than those observed in thin-film nanostripes with periodically modulated width (~10 GHz).[16,17] This discrepancy can be attributed to the different physical origin of collective spin excitations. In the former case, one starts from a set of discrete frequencies (corresponding to individual resonances inside each stripe) whose degeneration in energy is removed by dynamical dipolar coupling which is a relatively week perturbation of the individual resonances. In the latter case, one starts from the continuous frequency dispersion of spin waves in nanowaveguide with straight edges and forbidden bands appear as the result of a relatively weak Bragg backscattering at the edge steps of the width-modulated nanostripe.

**4.2 Arrays of alternated Co/NiFe stripes**

Another example of 1D magnonic crystal is the periodic arrays of alternating NiFe and Co stripes in direct contact fabricated by using the two-step fabrication method recalled at the end of Section 2. A SEM image of this sample, with lateral periodicity *a*=500 nm and stripe widths of 250 nm for both the Co and NiFe materials, is shown in Fig. 7.[12]

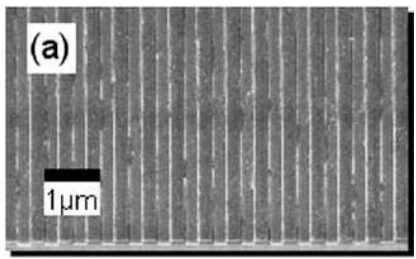

**Fig. 7** SEM image of a magnonic crystal in the form of a 1D periodic array of 30-nm-thick Permalloy and Cobalt nanostripes, each of width 250 nm. Reprinted with permission from [12].Z. K. Wang, V. L. Zhang, H. S. Lim, S. C. Ng, M. H. Kuok, S. Jain, and A. O. Adeyeye, Appl. Phys. Lett. **94**, 083112 (2009). © 2009, American Institute of Physics.

The group at University of Singapore exploited BLS to study spin waves propagation in such system. A relatively large bandwidth (about 2.5 GHz in width) was found for the fundamental magnonic band (see Fig.8). The observed Brillouin peaks were generally sharp, suggesting that the spin waves are weakly attenuated as they propagate through the magnonic crystal. Hence, any loss in devices based on this structure is expected to be small. In a previous theoretical calculation[41] it was pointed out that there are two different regimes of *dipole* coupling for such arrays. The occurrence for a particular regime depends on the location of a particular magnonic band with respect to the frequency band for the Damon-Eshbach[49] surface mode ("Damon-Eshbach band") for the materials. If a particular magnonic band is inside the Damon-Eshbach bands for both materials simultaneously, one can expect propagation of a travelling spin wave across the whole structure. This spin wave will represent the collective mode for the quasi-crystal which scatters on periodical inhomogeneities of the microwave magnetic susceptibility tensor,[51] the situation is very similar to magnonic crystals based on a periodical variation of the internal static field[52,53] or notched spin waveguides.[17] However, if a particular mode is outside the Damon-Eshbach band for one material, the collective mode will represent resonant precession in the form of a standing wave across the stripe for which this particular frequency is inside the Damon-Eshbach band. For the stripes made from the second material magnetisation precession will be forced and driven by the dipole fields from the resonant stripes. This driven magnetisation precession allows the collective mode to tunnel[54] between the resonating stripes. Tunnelling through a magnetic material is more efficient than through an air gap, as microwave susceptibility is larger for the former. This results in a stronger dipole coupling for the alternating contacting stripe array than for the arrays of stripes separated by air gaps.

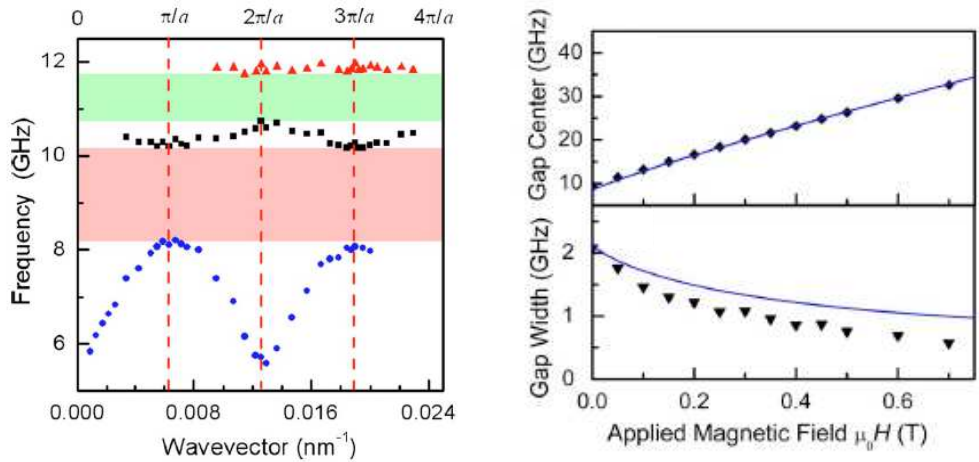

**Fig. 8** (Left panel) Frequency dispersion of collective modes measured by BLS measure in zero external applied field. Dashed lines indicate the BZ boundaries. (Right panel) Magnetic field dependence of the frequency centre and width of the first band gap on applied magnetic field. Reprinted with permission from [12]. Z. K. Wang, V. L. Zhang, H. S. Lim, S. C. Ng, M. H. Kuok, S. Jain, and A. O. Adeyeye, Appl. Phys. Lett. 2009 **94** 083112. © 2009, American Institute of Physics.

As followed from the material parameters given in Ref. 12, the Damon-Eshbach bands for the two materials do not overlap. A simulation based on the theory described in Ref.[41] shows that the dynamic for this structure is characterised by two families of collective modes separated in frequency. The lower-frequency family ("Permalloy family") is characterized by resonant precession of magnetisation in Permalloy stripes and a forced magnetisation motion in the Cobalt stripes. For the higher-frequency family the magnetic dynamics is opposite. It is resonant in Cobalt and forced in Permalloy. Obviously, the fundamental mode for this structure belongs to the Permalloy family and is characterised by a resonant precession of magnetisation in Permalloy stripes. Precession in all Permalloy stripes is strongly correlated (phase locked) by dipole fields which are "amplified" (with respect to the air gaps) by the high microwave magnetic permittivity material filling in the gaps between the resonating stripes. In the same work,[12] the field dependence of the band structure under the application of the external magnetic field was also studied. They found that the entire band structure is shifted up in frequency for increasing the applied field H strength while both the band gaps become narrower. These measurements indicate that the magnonic crystal exhibits band gap tunability, an important property which could find applications in the control of the generation and propagation of information-carrying spin waves in devices based on these crystals.

**5. 1 Experimental study of 2D magnonic crystals: Squared antidot array**

Different from the case of 1D magnonic crystals consisting of infinite stripes with uniform magnetisation, the study of 2D MC has to face the problem that pronounced magnetisation inhomogeneity is unavoidable. This makes it impossible to use a full-analytical theoretical approach and pushes towards the use of a micromagnetic approach. Here we first review the properties of a large-area $Ni_{80}Fe_{20}$ squared antidot array, whose SEM image is shown in Fig. 9, fabricated using the method described in Section2. The final structure consisted of circular holes with diameter of 250 nm and center-to-center spacing of 400 nm embedded into a 30 nm thick film.[55] In Fig. 9, we report the frequency dispersion measured by BLS in the Voigt geometry (magnetic field perpendicular to the transferred wave vector in the scattering process, $k$) when the external field is applied either at $\varphi = 45°$ or $\varphi = 0°$. It can be seen that for $\varphi = 45°$ all the detected modes are dispersion-less each with a fixed frequency in the whole range of investigated values of $k$. The spatial profiles of such modes, calculated by micromagnetic simulations[56] and reported in Fig. 10, indicate that these modes have a stationary character resonating in the antidot regions between neighboring holes. In particular, the 45deg-1 and 45deg-4 deg modes, are localized close to the holes edges where the local magnetisation is perpendicular and parallel to the holed edges, respectively. The absence of a dispersive character of the modes can be understood considering the alternating and interlaced rows of holes which excludes the presence of continuous film portions (infinitely extended) where mode propagation could occur.

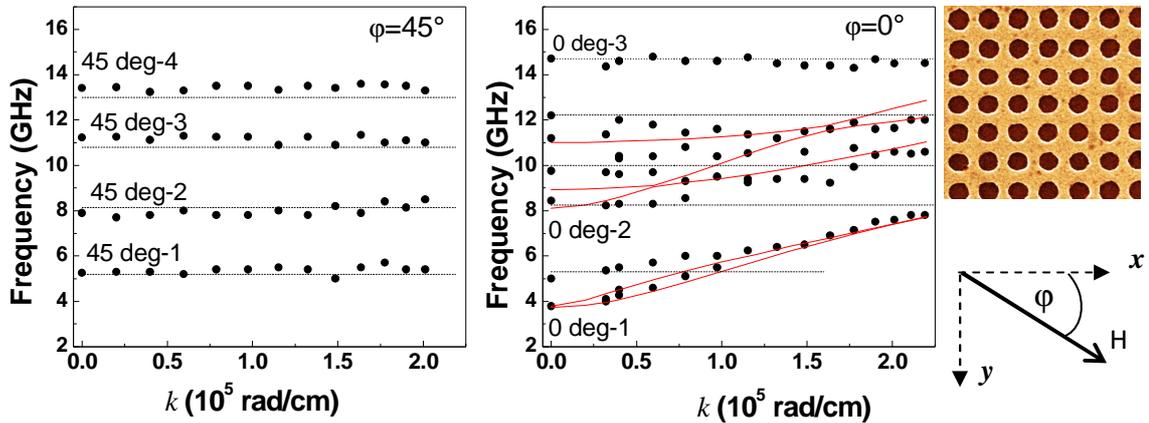

**Fig. 9** Spin wave dispersions for H = 1.0 kOe mT applied at $\varphi = 45°$ and $\varphi = 0°$. Points are the experimental data, while the horizontal dashed lines indicate the frequencies obtained by micromagnetic simulations for a transferred wave vector of $k = 0$. The full lines display calculated dispersions taken from Ref. [57]. Insets show the SEM image of the antidot array and the coordinate system. Reprinted with permission from [57]. Silvia Tacchi, Marco Madami, Gianluca Gubbiotti, Giovanni Carlotti, Adekunle O. Adeyeye, Sebastian Neusser, Bernhard Botters and Dirk Grundler 2010 IEEE Trans. on Magn. **46**, 172. (© 2010 IEEE) Reprinted with permission from [57]. M. Kostylev, G. Gubbiotti, G. Carlotti, G. Socino, S. Tacchi, C. Wang, N. Singh, A. O. Adeyeye and R. L. Stamps 2008 *J. Appl. Phys*. 103 07C507. © 2008, American Institute of Physics.

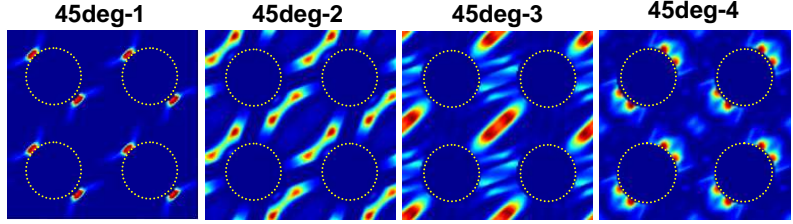

Fig. 10. Spatial distribution of spin precession amplitudes for an applied field H=1.0 kOe at φ= 45°. The numbers 1 to 4 refer to different eigenfrequencies labeled in Fig. 10(a). Reprinted with permission from [57]. Silvia Tacchi, Marco Madami, Gianluca Gubbiotti, Giovanni Carlotti, Adekunle O. Adeyeye, Sebastian Neusser, Bernhard Botters and Dirk Grundler 2010 IEEE Trans. on Magn. 46, 172 . (© 2010 IEEE)

Conversely, for φ = 0°, some of the modes exhibit a clear frequency dispersion with positive slope (positive group velocity) as a function of *k*. This is true, in particular, for the mode at low frequency (0deg-1 and 0deg-2), which have been identified as edge modes in Fig. 11. In addition, micromagnetic simulations indicate that these modes extend along the effective stripe perpendicular to the field direction, because of the interconnection characterized by a relatively large spin precession amplitude between nearest neighbouring holes in *y* direction. The simulations thereby confirm the hypothesis on the extended-mode character advanced in a preliminary work where a semi-quantitative analytical model was exploited to identify the character of dispersive modes.[57] Here, the antidot array was modelled as constituted by two families of orthogonal stripes of effective width equal to the interholes distance (*w* = 150 nm). Effectively the stripe forms a spin-wave waveguide.[56] Eventually, the 0deg-3 mode measured at about 14.7 GHz has the character of a standing spin wave resonating across the region between nearest holes where magnetisation is parallel to the magnetic field.

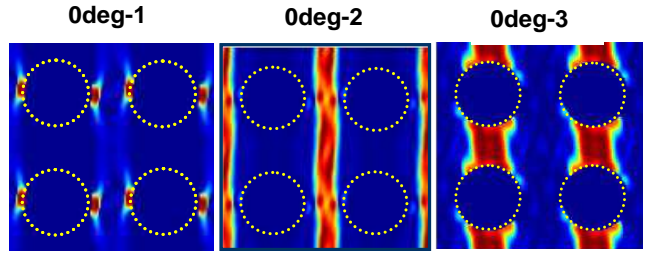

Fig. 11. Spatial distribution of spin precession amplitudes for an applied field H=1.0 kOe at φ = 0°. The numbers 1 to 3 refer to different eigen-frequencies labelled in Fig. 10 (b). Reprinted with permission from [55]. Silvia Tacchi, Marco Madami, Gianluca Gubbiotti, Giovanni Carlotti, Adekunle O. Adeyeye, Sebastian Neusser, Bernhard Botters and Dirk Grundler 2010 IEEE Trans. on Magn. **46**, 172. (© 2010 IEEE)

With these experiments we show that an antidot lattices support both propagating and resonating standing spin waves depending on the relative orientation of the external field with respect to the lattice mesh. Therefore, this system can be used to control and modify the spin wave spectrum by changing the holes dimension, distance and the symmetry of the lattice as well as the direction of the external applied field.[58]

### 5.2 2D planar Magnonic crystals: Arrays of circular dots

The first attempts to prove the effect of dynamical coupling on the spin excitations of 2D arrays of dots were made by studying the effect of interdot distance on the magnetic modes frequency.[59,60,61]

BLS experiments carried out on squared arrays of micron-size circular NiFe dots reveal the presence of a fourfold anisotropy, attributed to the dynamic magnetic dipole interaction between the dots at small inter-element distances. However, in these pioneering experiments neither dispersive modes nor bands formation have been detected. Furthermore, no quantitative description of the interplay between static and dynamic stray fields was provided.

More recently, a systematic study of the spin-wave frequency has been performed in squared arrays of polycrystalline $Ni_{80}Fe_{20}$ circular dots with radius $R$=100 nm, thickness $L$=50 nm, and variable spacing $\Delta$=50, 100, 200, 400, and 800 nm. These arrays of dots were fabricated by a combination of *e*-beam lithography, *e*-beam evaporation, and lift-off processes at Kyoto University. Figure 12 reports the measured frequency by BLS for arrays of disks with different spacing recorded for an external field H=2.0 kOe, applied along the [10] lattice direction, and an incidence angle of light θ =10°.[38,62] In the same figure we show calculated frequency bands obtained by a full analytical theory for collective modes in 2D arrays of disks.[63] The comparison of the measurements with the calculated modes spatial distribution allowed us to assign the modes character. Starting from the lowest frequency, we recognize a laterally localized mode, or end-mode of zero order, (0-EM), a backward-like mode (i.e. with oscillations along the direction of the applied field, 2-BA), the fundamental or quasi-uniform mode (F) and three Damon-Eshbach like modes (i.e. with oscillations perpendicular to the applied field,1-DE, 2-DE, and 3-DE). The modes nomenclature is consistent with the one introduced for the modes of non-interacting magnetic dots.[36] The labels BA and DE has been adopted for historical reasons from the film waves,[64] although in the case of dots we deal with standing waves, rather than with travelling waves. For large interdot separation, each mode is characterized by a single frequency: in this limit the coupling is negligible, and the eigenfrequencies do not depend on $k$. From the analysis of the measured spectra, we notice that, as soon as the disks separation decreases below 200 nm the 0-EM and the 2-BA modes suffer a large frequency increment with respect to the values (5.6 GHz and 7.6 GHz) they have for large interdot distance ($\Delta$=800nm). Micromagnetic simulations[38,63] indicate that these modes retain their own character being highly localized close to the disks edges in the direction of the external applied field. Their frequency increase is mainly a static effect, associated to the variation of the internal magnetic field inside the disks when they becomes very close one to another, rather than a consequence of the dynamical coupling between standing modes of adjacent disks. At the same time, the modes at higher frequencies classified as 1, 2 and 3-DE modes, are almost insensitive to the inter-disks spacing reduction having almost the same frequency as a function of the spacing.

The most pronounced changes associated to dynamical coupling between disks are observed in the measured spectra for the quasi-fundamental mode (F) which exists in the central portion of the disks and which splits into three modes having different profiles and which can be either at larger frequency or at lower frequency with respect to the original frequency (11.2 GHz for $\Delta$=800 nm). These three modes appear at 10.8, 11.8, and 13 GHz for $\Delta$ =50 nm. Full analytical calculations indicate that, when s becomes comparable to the dot diameter, the interdot coupling gives rise to the appearance of bands; within each band the frequency of the collective modes depends on $k$. The upper and lower limits of the calculated bands correspond to an out-of-phase ($k=k_{BZ}$, boundary of the BZ) and in-phase ($k$=0 centre of the BZ) oscillation of dynamical magnetisations in coupled dots, respectively. Calculation also reproduce the frequency increase of the measured band centre and width change as $\Delta$ is decreased. The F mode feels the strongest interdot coupling as already described for dipolarly coupled stripes. The strongest coupling of the F can be traced to its non-vanishing average magnetisation: in this case the coupling between adjacent dots is relatively large, thanks to the dipolar field as shown in Fig. 2. The interdot coupling for the other modes, all having a vanishing or small average magnetisation, is weaker and mainly due to the non-uniformity of the dynamic magnetisation.

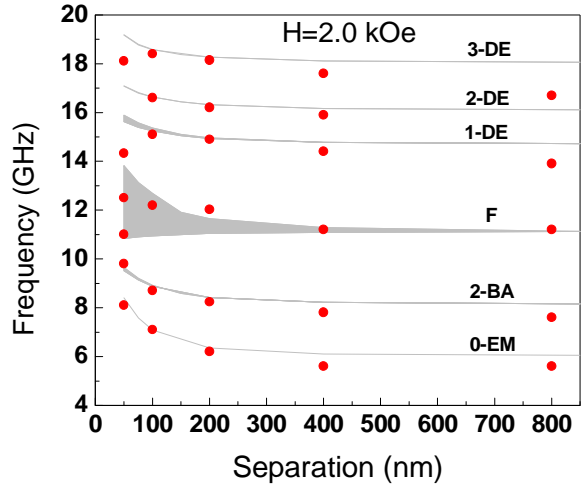

**Fig. 12** Comparison between the measured (red dots) and calculated frequency dependence of the collective spin modes on the interdot separation Δ. At small Δ the modes form bands, represented by gray areas in the figure. Reprinted with permission from [63]. L. Giovannini, F. Montoncello, and F. Nizzoli 2007 *Phys. Rev. B* **75** 024416. © 2007, American Institute of Physics.

## 6. Conclusions

In this article we reviewed the main experimental results, appeared in the literature during the last few years, concerning thermal collective spin wave excitations in 1D and 2D planar magnonic crystals, carried out by the Brillouin light scattering technique. Arrays of antidots and interacting magnetic elements, such as stripes and dots, constitute model systems which have been used to learn a lot of basic properties of spin-waves in this new kind of artificial materials. Band-gap tunability has been demonstrated, acting on either the geometrical and material parameters, as well as on the applied magnetic field. An important parameter which needs to be controlled in view of possible application of MC in commercial devices is the damping of the collective excitations on the array which limits the interaction to take place only among a finite number of elements. Efforts should be devoted to find new materials characterized by an high-saturation magnetisation and a low damping. This combination will maximize both group velocity and free propagation path of spin waves. New features are expected to appear when the repetition unit cell contains two or more elements with different shapes and materials. In terms of dynamical properties, one could try to put in evidence the possible existence of modes with in-phase and out-of-phase precession of the magnetisation in the unit cell, as it happens for "acoustic" and "optical" modes in the phonon spectrum of crystals with a complex base. From a technological point of view, a proper understanding of the magnetisation dynamics in MC underpins the operation of microwave signal processing and application in magnetic logic devices able to work in the GHz frequency range, where the wavelengths of magnonic excitations are shorter than those of light. Hence, magnonic meta-materials offer better prospects for miniaturisation at these frequencies. Moreover, in magnetic storage media, magnetic nano-structures have already been combined with nano-electronics (e.g. in read heads and spin torque magnetic random access memories) and optics (e.g. in magneto-optical disks). Therefore, filters and logic gates based on MC with a nanoscale period situated between spin wave emitter and detector are likely to be exploited in the near future. The main requirement will be the presence of a magnetic field controlled magnonic band gap within the frequency of the spin wave emitter and detector.

**Aknowledgments**

Authors acknowledge the European Community's Seventh Framework Programme (FP7/2007-2013) under Grant Agreement n°228673 (MAGNONICS), the Ministero per l'Università e la Ricerca under PRIN-2007 project (Prot. Grant No. 2007X3Y2Y2), funding from the Australian Research Council and the Ministry of Education, Singapore (Grant No. R-263-000-437-112).